\documentclass[a4paper]{article}
\usepackage{amssymb}
\usepackage{graphicx}
\usepackage{algorithm}
\usepackage{algorithmic}
\usepackage{diagbox}
\usepackage{color}
\usepackage{amsmath}
\usepackage{multirow}
\usepackage{array}

\usepackage{INTERSPEECH2020}

\title{Advancing Multiple Instance Learning with Attention Modeling for Categorical Speech Emotion Recognition}
\name{Shuiyang Mao\textsuperscript{1}, P. C. Ching\textsuperscript{1}, C.-C. Jay Kuo\textsuperscript{2} and Tan Lee\textsuperscript{1}}
\address{\textsuperscript{1}Department of Electronic Engineering, The Chinese University of Hong Kong, Hong Kong\\
\textsuperscript{2}Ming Heish Department of Electrical Engineering, University of Southern California, USA}
\email{maoshuiyang@link.cuhk.edu.hk, \{pcching, tanlee\}@ee.cuhk.edu.hk, cckuo@sipi.usc.edu}

\begin{document}
%
\maketitle
\begin{abstract}
Categorical speech emotion recognition is typically performed as a sequence-to-label problem, i.\,e., to determine the discrete emotion label of the input utterance as a whole. One of the main challenges in practice is that most of the existing emotion corpora do not give ground truth labels for each segment; instead, we only have labels for whole utterances. To extract segment-level emotional information from such weakly labeled emotion corpora, we propose using multiple instance learning (MIL) to learn segment embeddings in a weakly supervised manner. Also, for a sufficiently long utterance, not all of the segments contain relevant emotional information. In this regard, three attention-based neural network models are then applied to the learned segment embeddings to attend the most salient part of a speech utterance. Experiments on the CASIA corpus and the IEMOCAP database show better or highly competitive results than other state-of-the-art approaches.
\end{abstract}
\noindent\textbf{Index Terms}: categorical speech emotion recognition, weak labeling, multiple instance learning, attention modeling
\section{Introduction}
\label{sec:intro}

Automatic speech emotion recognition (ASER) aims to decode emotional content from audio signals. It has constituted an active research topic in the field of human-machine interaction (HCI). Detection of lies, monitoring of call centers, and medical diagnoses are also considered as promising application scenarios of speech emotion recognition.

Categorical speech emotion recognition at utterance level can be formulated as a sequence-to-label problem. The input utterance is divided into a sequence of acoustic segments, and the output is a single label of emotion type. A few previous studies explored the use of segment units for categorical speech emotion recognition and demonstrated that combining segment-level prediction results led to superior performance \cite{jeon2011sentence, shami2005segment, schuller2006timing}. These prior works were mostly based on conventional models, such as \textit{support vector machine} (SVM) and \textit{K-nearest neighbors} (K-NN), for segment classification. In this paper, a \textit{convolutional neural network} (CNN) is applied to extract emotionally relevant features, i.\,e., to learn emotion relevant segment embeddings. Also, most of the existing emotion corpora do not provide ground truth labels at segment level; instead, they only have labels for whole utterances. A viable solution is to learn local concepts from global annotations, which is the main idea of \textit{multiple instance learning} (MIL) \cite{amores2013multiple, xu2014deep}. MIL has been successfully applied to sound event detection (SED) \cite{su2017weakly, kumar2016audio}, speech recognition \cite{wang2018comparing}, and image analysis \cite{xu2014deep}. In the MIL problem statement, the training set contains labeled bags that comprise many unlabeled instances, and the task is to predict the labels of unseen bags and instances. For categorical speech emotion recognition, each utterance is treated as a bag, and segments within the utterance as instances. One main feature of this work is the application of deep learning of feature representation in the MIL framework to learn segment embeddings from weak labels at utterance level. Compared to the raw features such as MFCC, energy, or pitch, the learned segment embeddings are more tied to the task of interest, thus naturally highlighting salient portions of the data, which we conjecture would offer an advantage in the final classification.

A key question is how to enable a deep learning model to identify and focus on the most salient parts of a speech utterance when making an utterance-level decision with the learned segment embeddings. In this regard, attention neural network models are investigated. The key idea behind the attention mechanism is to align the input-output sequences such that, in the decoding phase, the major contribution is from the corresponding encoded information. In contrast, the effect of irrelevant ones is minimized. In this work, attention modeling is expected to facilitate a structurally meaningful composition of the utterance representation from the learned emotionally relevant segment embeddings. This is the first attempt to combine MIL-based deep learning of segment embedding with attention modeling for categorical speech emotion recognition.
\begin{figure}[t]
  \centering
  \includegraphics[width=\linewidth]{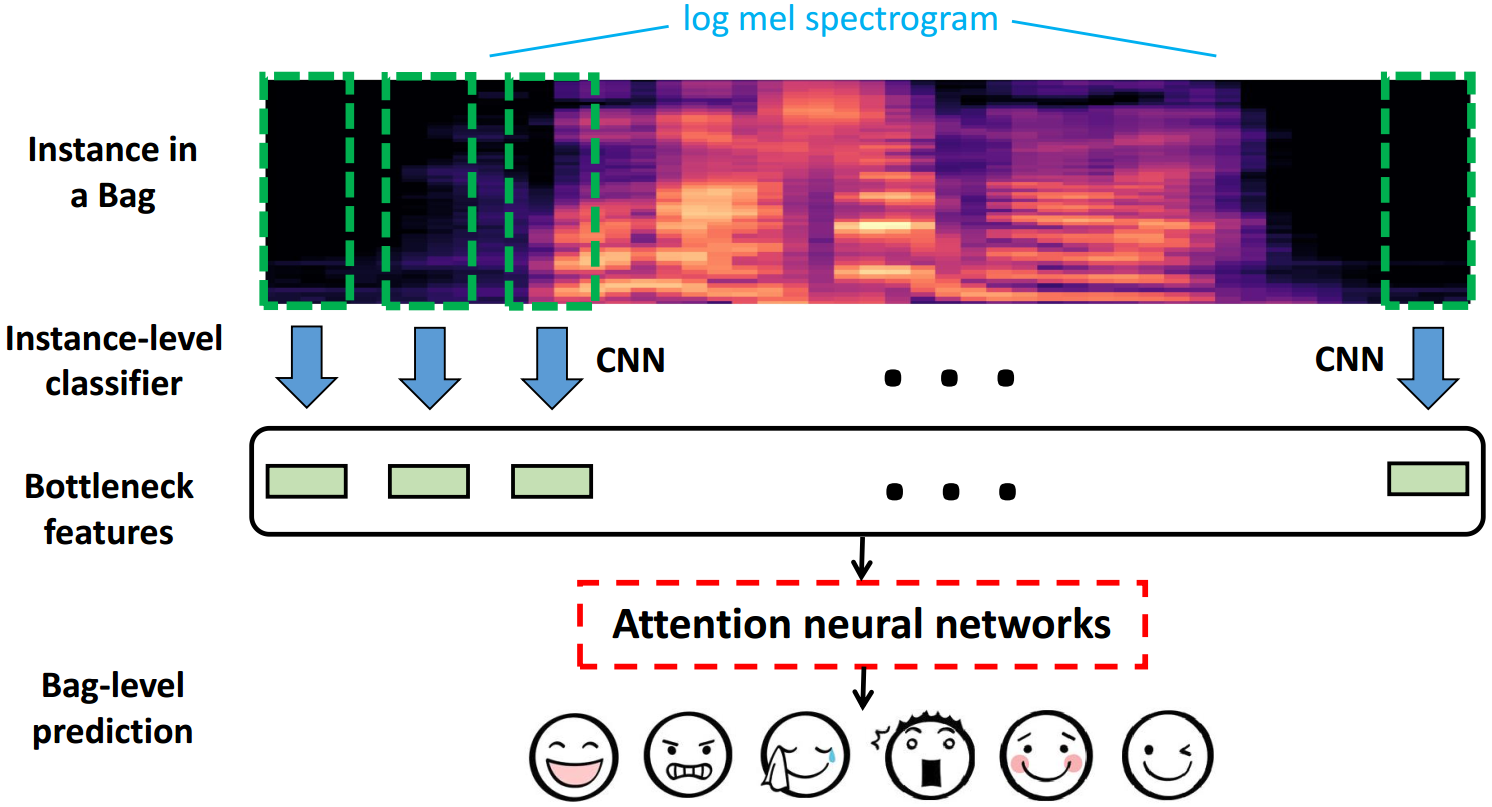}
  \caption{Illustration of the proposed framework for categorical speech emotion recognition.}
  \label{fig:system}
\end{figure}
\section{Methodology}
Figure~\ref{fig:system} illustrates a schematic approach of the proposed method. It comprises a CNN model trained to learn emotionally salient segment embeddings from log-Mel filterbank features of individual segments. The learned segment embeddings are then used as inputs for utterance-level emotion recognition, which is achieved by a dense-layer neural network implemented with various attention mechanisms.
\subsection{MIL-based segment embedding learning}
We formulate our segment-based approach as a MIL problem following the instance space paradigm \cite{amores2013multiple}. Each utterance (bag) is first divided into a sequence of segments (instances). These individual segments are then used to train a CNN model. The learned CNN aims to generate emotionally salient embeddings for each segment.
\begin{figure}[t]
  \centering
  \includegraphics[width=\linewidth]{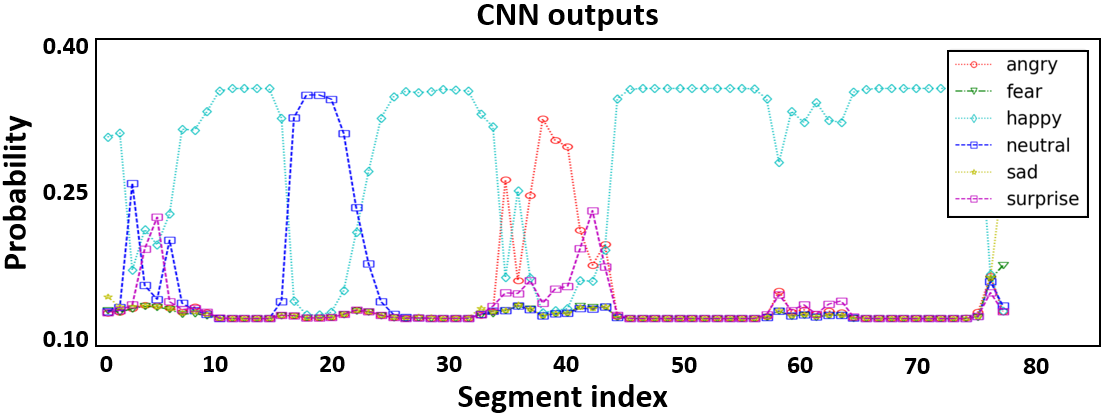}
  \caption{An example of CNN outputs for the audio file ``Happy\_liuchanhg\_382.wav" from the CASIA corpus.}
  \label{fig:prop}
\end{figure}
\subsubsection{Log Mel spectrograms}

For the segment-level features, we use the 64-bin log Mel filterbanks, which have been extensively evaluated in the existing literature \cite{huang2014speech, satt2017efficient, zhang2018convolutional}. They are computed by \textit{short-time Fourier transform} (STFT) with a window length of 25 ms, hop length of 10 ms, and FFT length of 512. Subsequently, 64-bin log Mel filterbanks are derived from each short-time frame, and the frame-level features are combined to form a time-frequency matrix representation of the segment.

\subsubsection{Segment embeddings}

Our segment-based method must address how to train a segment-level model without access to a training set of labeled segments. To address this problem, we follow the  most straightforward approach, called Single Instance Learning (SIL) \cite{bunescu2007multiple}, i.e., each segment inherits the label of the utterance where it lies. A CNN is then trained on the resulting dataset. The outputs of the penultimate layer of the trained CNN, which we refer to as \textit{segment embeddings} in this work, are stored and will be employed as inputs to the subsequent recognition part. Besides, a softmax layer sits on top of the CNN model and aims to predict a probability distribution \(\textbf{P}\) as follows:
\begin{align}
  \textbf{P} = [p(e_1),\ p(e_2),\ \cdots,\ p(e_K)]^T
  \label{eq1}
\end{align}
where \(K\) denotes the number of possible emotions. 

Figure~\ref{fig:prop} shows an example of a probability distribution predicted by the trained CNN for the audio file ``Happy\_liuchanhg\_382.wav" from the CASIA corpus. It can be observed that: (1) the probability distribution of each segment changes across the whole utterance; (2) most of the segments convey information that conforms to the utterance where they lie; and (3) there are segments within one utterance that do not convey any information about the target emotion class or that are more related to other classes, which constitutes confusing information. If we can place additional focus on these more relevant segments, system performance might be improved. In this regard, we have developed three attention-based neural networks, which are described in detail in the following section.

\subsection{Attention-based neural networks}

\begin{figure*}[thbp]
  \centering
  \includegraphics[width=0.66\linewidth]{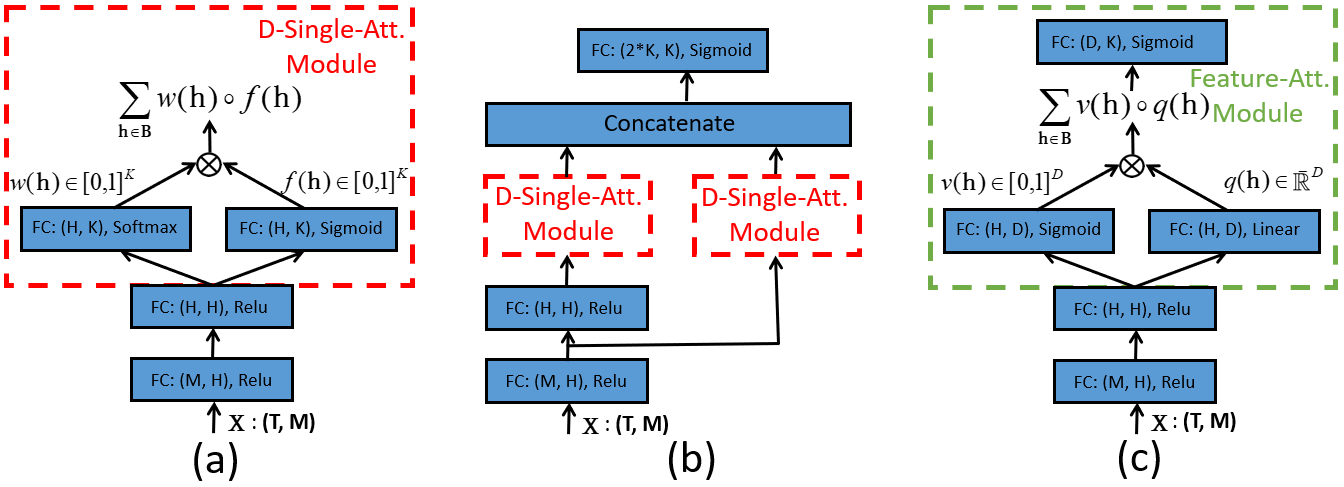}
  \caption{(a) Decision-level single attention neural network; (b) decision-level multiple attention neural network; (c) feature-level attention neural network. (\(\circ\): Hadamard product; \(\sum\): element-wise summation; T: length of input sequence; M: dimension of input bottleneck features; H: dimension of FC layer; K: number of emotion classes; D: dimension of feature-level attention function.)}
  \label{fig:attention}
\end{figure*}

Attention neural networks assume that the bag-level prediction can be constructed as a weighted sum of the instance-level predictions. Herein, three attention-based neural network models are investigated and compared, i.e., decision-level single attention (D-Single-Att.) \cite{kong2018audio}, decision-level multiple attention (D-Multi-Att.) \cite{yu2018multi} and feature-level attention (Feature-Att.) \cite{kong2019weakly}, as shown in Figure~\ref{fig:attention}(a)-(c), respectively. We denote the input segment embeddings within a certain speech utterance as \(\textbf{X} \in \mathbb{R}^{T \times M}\), where \(T\) is the number of segments and \(M\) represents the dimension of segment embeddings. The output of the second fully-connected (FC) layer is denoted as \(\textbf{h}\), which has a dimension of \(120\), i.\,e., H is set to 120 in Figure~\ref{fig:attention}.
\subsubsection{Decision-level single attention}

In the decision-level single attention model (as shown in Figure~\ref{fig:attention}(a)), an attention function is applied on the predictions of the instances to obtain the bag-level prediction:
\begin{align}
  F(\textbf{B})_k = \sum\limits_{\textbf{h} \in \textbf{B}} w(\textbf{h})_k f(\textbf{h})_k
  \label{eq2}
\end{align}
where \(k\) denotes the \(k\)-th emotion class of the instance-level prediction \(f(\textbf{h}) \in [0,1]^{K}\) and the bag-level prediction \(F(\textbf{B}) \in [0,1]^{K}\), and \(w(\textbf{h})_k \in [0,1]\) is a weight of \(f(\textbf{h})_k\) that we refer to as a \textit{decision-level attention function}:
\begin{align}
  w(\textbf{h})_k = s(\textbf{h})_k / \sum\limits_{\textbf{h} \in \textbf{B}} s(\textbf{h})_k
  \label{eq3}
\end{align}
where \(s(.)\) can be any non-negative function (i.e., Softmax nonlinearity) to ensure that attention \(w(.)\) is normalized. Both the attention function \(w(.)\) and the instance-level classifier \(f(.)\) depend on a set of learnable parameters.

\subsubsection{Decision-level multiple attention}

The decision-level multiple attention model is an extension of the above decision-level single attention model. It consists of several single attention modules (we herein use two attention modules, as shown in Figure~\ref{fig:attention}(b)) applied to intermediate neural network layers. The outputs of these attention modules are concatenated.

\subsubsection{Feature-level attention}

The limitation of the above decision-level attention neural networks is that the attention function \(w(.)\) is only applied to the prediction of the instances \(f(\textbf{h})\). To address this constraint, we also investigate the effect of applying the attention function to the outputs of the hidden layers, which we refer to as the feature-level attention (as shown in Figure~\ref{fig:attention}(c)), in which the bag-level representation \(\textbf{U}\) can be modeled as: 
\begin{align}
  \textbf{U}_d = \sum\limits_{\textbf{h} \in \textbf{B}} v(\textbf{h})_d q(\textbf{h})_d
  \label{eq4}
\end{align}
where \(d\) denotes the \(d\)-th dimension of the hidden layer output \(q(\textbf{h}) \in \mathbb{R}^{D}\) and the bag-level representation \(\textbf{U} \in \mathbb{R}^{D}\); and \(v(\textbf{h})_d \in [0,1]\) is a weight of \(q(\textbf{h})_d\) that we refer to as a \textit{feature-level attention function}:
\begin{align}
  v(\textbf{h})_d = u(\textbf{h})_d / \sum\limits_{\textbf{h} \in \textbf{B}} u(\textbf{h})_d
  \label{eq5}
\end{align}
where \(u(.)\) can be any non-negative function (i.e., Sigmoid nonlinearity) to ensure that attention \(v(.)\) is normalized. Both the attention function \(v(.)\) and the instance-level feature mapping function \(q(.)\) depend on a set of learnable parameters. The prediction of a bag \(\textbf{B}\) can then be obtained by classifying the  bag-level representation \(\textbf{U}\) as follows:
\begin{align}
  F(\textbf{B}) = g(\textbf{U})
  \label{eq6}
\end{align}
where \(g(.)\) is the final classifier that corresponds to the last neural network layer. 

\section{Emotion Corpora}
Two different emotion corpora are used to evaluate the validity of the proposed method, namely, a Chinese emotional corpus (CASIA) \cite{tao2008design} and an English emotional database (IEMOCAP) \cite{busso2008iemocap}, which have been extensively evaluated in the literature.

Specifically, the CASIA corpus \cite{tao2008design} contains 9,600 utterances that are simulated by four subjects (two males and two females) in six different emotional states, i.\,e., angry, fear, happy, neutral, sad, and surprise. In our experiments, we only use 7,200 utterances that correspond to 300 linguistically neutral sentences with the same statements. All of the emotion categories are selected.

The IEMOCAP database \cite{busso2008iemocap} was collected using motion capture and audio/video recording over five dyadic sessions with 10 subjects. At least three evaluators annotated each utterance in the database with the categorical emotion labels chosen from the set: angry, disgusting, excited, fear, frustrate, happy, neutral, sad, surprise, and others. We consider only the utterances with majority agreement (i.\,e., at least two out of three evaluators assigned the same emotion label) over the emotion classes of angry, happy (combined with the ``excited" category), neutral and sad, which results in 5,531 utterances in total.

\section{Experiments}
\label{sec:experiments}

\subsection{Setup}
In our experiment, the size of each speech segment is set to 32 frames, i.\,e., the total length of a segment is 10 ms \(\times\) 32 + (25 - 10) ms = 335 ms, shifting 60 ms each time. In this way, we collected approximately 200,000 segments for the CASIA corpus and 300,000 segments for the IEMOCAP database, respectively. Moreover, since the input length for our attention neural networks has to be equal for all samples, we heuristically set the maximal length for each speech utterance to the average duration of each dataset, i.\,e., 2.07 s for CASIA and 4.55 s for IEMOCAP, respectively. Longer speech utterances are cut at the maximal length, and shorter ones are padded with zeros.

The architecture of the CNN model is similar to the \textit{SegCNN} model as used in our previous work  \cite{mao2019deep}. The only change we made was to the last three FC layers, i.\,e., \(\{256, 64, K\}\) units, respectively, where \(64\) and \(K\) correspond to the dimension of segment embeddings and the number of possible emotions, respectively. In the training stage, for both CNN and attention neural networks, ADAM \cite{kingma2014adam} optimizer with the default setting in Tensorflow \cite{abadi2016tensorflow} was used, with an initial learning rate of \(0.001\) and an exponential decay scheme with a rate of \(0.8\) every two epochs. The batch size was set to \(128\). Early stopping with patience of \(3\) epochs was utilized to mitigate an overfitting problem. 

\begin{figure*}[thbp]
  \centering
  \includegraphics[width=0.816\linewidth]{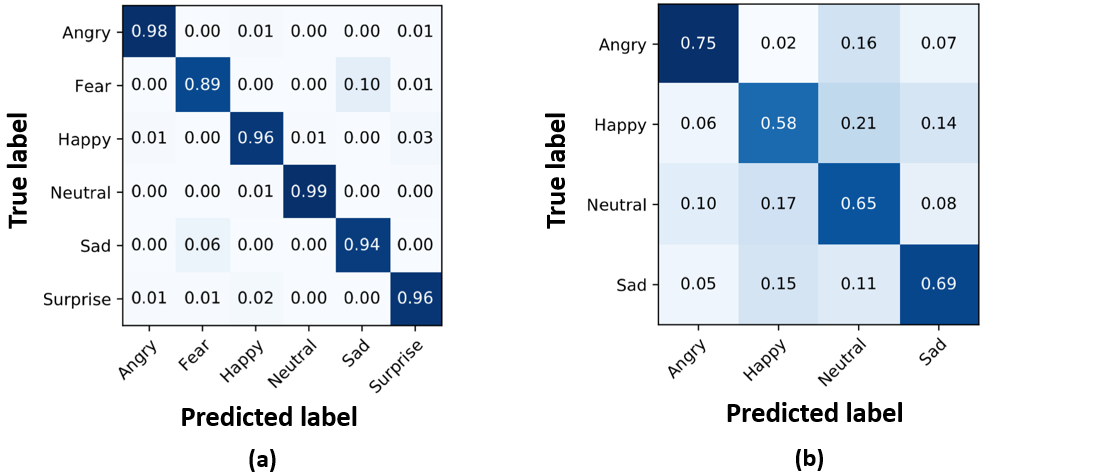}
  \caption{Confusion matrices obtained using feature-level attention modeling for: (a) CASIA corpus and (b) IEMOCAP database.}
  \label{fig:confusionM}
\end{figure*}

For the CASIA corpus, we perform leave-one-fold-out ten-fold cross-validation experiments. For the IEMOCAP database, the leave-one-session-out five-fold cross-validation method is carried out. For both datasets, a second cross-validation is performed since we need to utilize the segment-level results to train the attention neural networks. The results are presented in terms of unweighted accuracy (UA).

\subsection{Baseline systems}

\subsubsection{Maxout segment embedding (CNN-MAX-RF)}
In the CNN-MAX-RF baseline, the maxout pooling is directly applied on the segment embedding \(\textbf{X} \in \mathbb{R}^{T \times M}\):
\begin{align}
  \textbf{U}_m = \mathop{\textnormal{max}}\limits_{0\leq t \leq T}\ \{\textbf{X}_m\}
  \label{eq2}
\end{align}
A \textit{Random Forest} (RF) is then used to make the utterance-level prediction based on the resultant utterance representation \(\bf U\). 

\subsubsection{Averaging segment embedding (CNN-AVG-RF)}
The CNN-AVG-RF baseline is similar to the above CNN-MAX-RF baseline. The only difference is that the maxout pooling in the CNN-MAX-RF baseline is replaced by an average pooling in the CNN-AVG-RF baseline.

\subsubsection{Maxout pooling (CNN-MP)}
In the CNN-MP baseline, the maxout pooling is applied on the instance-level prediction \(f(\textbf{h}) \in [0,1]^{K}\) across a certain speech utterance to obtain the bag-level prediction:
\begin{align}
  F(\textbf{B})_k = \mathop{\textnormal{max}}\limits_{\textbf{h} \in \textbf{B}}\ \{f(\textbf{h})_k\}
  \label{eq2}
\end{align}

\subsubsection{Average pooling (CNN-AP)}
Similarly, for the CNN-AP baseline, the average pooling is applied on the instance-level prediction \(f(\textbf{h}) \in [0,1]^{K}\) across a particular speech utterance to obtain the bag-level prediction:
\begin{align}
  F(\textbf{B})_k = \mathop{\textnormal{mean}}\limits_{\textbf{h} \in \textbf{B}}\ \{f(\textbf{h})_k\}
  \label{eq2}
\end{align}
\begin{table}[t]
\renewcommand\arraystretch{1}
\caption{Comparison of UAs on the CASIA corpus.}
\label{tab:casia}
\centering
\resizebox{0.66 \linewidth}{!}{%
\begin{tabular}{lc}
\toprule
\bf{Methods for comparison} & \bf{UA [$\%$]} \\
\midrule                            
ELM-Decision Tree \cite{liu2018speech} & $89.60$ \\
DNN-HMM \cite{mao2019revisiting}  & $91.32$ \\
LSTM-TF-Att. \cite{xie2019speech}& $92.80$ \\
\midrule
CNN-MAX-RF (\textit{baseline}) & $91.10$\\
CNN-AVG-RF (\textit{baseline}) & $91.75$\\
CNN-MP (\textit{baseline}) & $92.11$\\
CNN-AP (\textit{baseline}) & $92.36$\\
\midrule
CNN-D-Single-Att. (\textit{ours})  & $93.75$\\
CNN-D-Multi-Att. (\textit{ours})  & $94.57$\\
CNN-Feature-Att. (\textit{ours})  & $\bf95.32$ \\
\bottomrule
\end{tabular}
}
\end{table}

~\\[-1.1cm]
\subsection{Results and analysis}

Tables 1-2 show the experimental results on the two mentioned emotion corpora, respectively. The following can be seen: (1) our baseline systems achieved respectable results on both datasets, which proved the effectiveness of the MIL-based framework; (2) the last two baselines (i.\,e., the CNN-MP baseline and the CNN-AP baseline) consistently outperformed the first two baselines (i.\,e., the CNN-MAX-RF baseline and the CNN-AVG-RF baseline). This performance gain might derive from the joint optimization of the aggregation strategy of segment-level representations and the utterance-level decision making of the last two baselines; (3) the attention-based methods substantially augmented the performance of the baseline systems overall. This is mainly attributed to the effectiveness of the attention modeling; (4) due to the positive combination of different attention modules, the decision-level multiple attention modeling achieved noticeably better performance than the decision-level single attention modeling on both datasets; (5) the feature-level attention modeling outperformed the decision-level attention neural networks by a significant margin. This is due to the fact that the dimension of \(v(\textbf{h})\) (i.e., \(D\)) can be any value, while the dimension of \(w(\textbf{h})\) is fixed to be the number of emotion classes \(K\). With an increase in the dimension of \(v(\textbf{h})\), the capacity of feature-level attention neural networks is increased; and (6) for the CASIA corpus, our feature-level attention-based system achieved the highest recognition accuracy of \(95.32\%\), establishing a new benchmark (to the best of our knowledge). For the IEMOCAP database, which might constitute a more challenging dataset, our methods also achieved competitive results. Figure~\ref{fig:confusionM} shows the confusion matrices obtained using feature-level attention modeling on both datasets, respectively.
\begin{table}[t]
\renewcommand\arraystretch{1}
\caption{Comparison of UAs on the IEMOCAP database.}
\label{tab:iemocap}
\centering
\resizebox{0.66 \linewidth}{!}{%
\begin{tabular}{lc}
\toprule
\bf{Methods for Comparison} & \bf{UA [$\%$]} \\
\midrule
CNN-LSTM \cite{satt2017efficient} & $59.40$ \\
DNN-HMM \cite{mao2019revisiting}  & $62.23$ \\
FCN-Att. \cite{zhang2018attention} & $63.90$\\
\midrule
CNN-MAX-RF (\textit{baseline}) & $61.08$\\
CNN-AVG-RF (\textit{baseline}) & $61.40$\\
CNN-MP (\textit{baseline}) & $62.11$\\
CNN-AP (\textit{baseline}) & $62.74$\\
\midrule
CNN-D-Single-Att. (\textit{ours})  & $63.34$ \\
CNN-D-Multi-Att. (\textit{ours})  & $65.82$ \\
CNN-Feature-Att. (\textit{ours})    & $\bf66.74$ \\
\bottomrule
\end{tabular}
}
\end{table}

\section{Conclusion}
\label{sec:conclusion}

In this paper, we proposed to combine multiple instance learning with attention neural networks for better modeling of categorical speech emotion recognition. Three attention-based neural network models were investigated and compared. Experimental results on two well-known emotion corpora showed competitive outcomes. Since we herein blindly used all segments to train the segment-level classifier, it is anticipated with proper segment selection strategy, better results are expected. More advanced neural network architectures and better algorithm optimization will also be investigated in the near future.


\newpage

\bibliographystyle{IEEEtran}
\begin{footnotesize}
\bibliography{mybib}
\end{footnotesize}

\end{document}